\documentclass{optica-article}

\journal{opticajournal} % for journals or Optica Open

\articletype{Research Article}

\begin{document}

\title{Magnetic-field-induced splitting of Rydberg Electromagnetically Induced Transparency (EIT) and Autler-Townes (AT) spectra in $^{87}$Rb vapor cell}

\author{Xinheng Li,\authormark{1,4} Yue Cui,\authormark{1,4}Jianhai Hao,\authormark{1} Fei Zhou,\authormark{1} Fengdong Jia,\authormark{1,*} Jian Zhang,\authormark{2} Feng Xie,\authormark{2} and Zhiping Zhong\authormark{1,3}}

\address{\authormark{1}School of Physical Sciences, University of Chinese Academy of Sciences, Beijing 100049, China\\
\authormark{2}Institute of Nuclear and New Energy Technology, Collaborative Innovation Center of Advanced Nuclear Energy Technology, Key Laboratory of Advanced Reactor Engineering and Safety of Ministry of Education, Tsinghua University, Beijing 100084, China\\
\authormark{3}CAS Center for Excellence in Topological Quantum Computation, University of Chinese Academy of Sciences, Beijing 100190, China \\
\authormark{4}The authors contributed equally to this work.}

\email{\authormark{*}fdjia@ucas.ac.cn} %% email address is required; see note below about the corresponding author designation

% use {asbstract*} to suppress the copyright line. Copyright information will be added in production

\begin{abstract*} 
We theoretically and experimentally investigate the Rydberg electromagnetically induced transparency (EIT) and Autler-Townes (AT) splitting of $^{87}$Rb vapor under the combined influence of a magnetic field and a microwave field. In the presence of static magnetic field, the effect of the microwave field leads to the dressing and splitting of each $m_F$ state, resulting in multiple spectral peaks in the EIT-AT spectrum. A simplified analytical formula was developed to explain the EIT-AT spectrum in a static magnetic field, and the calculations are in excellent agreement with experimental results.We further studied the enhancement of the Rydberg atom microwave electric field sensor performance by making use of the splitting interval between the two maximum absolute $m_F$ states under static magnetic field. The traceable measurement limit of weak electric field by EIT-AT splitting method was extended by an order of magnitude, which is promising for precise microwave electric field measurement.

\end{abstract*}

%%%%%%%%%%%%%%%%%%%%%%%%%%  body  %%%%%%%%%%%%%%%%%%%%%%%%%%
\section{Introduction}
Rydberg atom, highly excited atom with one or more electrons excited to states with high principal quantum number\cite{gallagher_1994}, exhibits unique properties including large radius, long lifetime, and high sensitivity to external fields\cite{Adams_2020}. These features make Rydberg atoms perfect candidate for precise measurements of external fields,which have important applications in determining the optical properties of materials and digital communication\cite{Sedlacek_2012,Song_19}. Experimentally realized in 2012 by Sedback \textit{et al.}\cite{Sedlacek_2012}, Rydberg atom-based microwave electric field measurement method primarily utilizes the Rydberg atoms’ high sensitivity to external fields and the phenomena of electromagnetically induced transparency (EIT) and AC Stark shift \cite{Fancher_2021,Fan_2015}. Specifically, EIT refers to the appearance of a narrow transparency peak in the absorption spectrum when a coupling light is applied to a three-level system\cite{JP_Marangos_1998}. Under the resonant dressing effect of the microwave field on Rydberg-to-Rydberg transitions, the AC Stark shift of Rydberg state levels leads to the symmetric splitting of EIT, known as Autler-Townes (AT) splitting. When scanning the frequency of the probe or coupling light, the interval of EIT-AT splitting has the following relationship with the microwave electric field strength\cite{Fancher_2021}:

\begin{equation}
\Delta f_p = \begin{cases}
\frac{\lambda_c}{\lambda_p}\frac{\Omega_{RF}}{2\pi} = \frac{\lambda_c}{\lambda_p}\frac{\mu}{h}E_{RF} ,    &  if\, \Delta_p \,is \,scanned \\
\frac{\Omega_{RF}}{2\pi} = \frac{\mu}{h}E_{RF} ,   &  if\, \Delta_c \,is \,scanned
\end{cases},
\end{equation}

where $h$ is the Planck constant, $\mu$ is the microwave transition dipole moment, $\lambda_p$ and $\lambda_c$ are the wavelengths of the probe light and coupling light, respectively. The spectral splitting interval $\Delta f_p$ is directly proportional to the Rabi frequency $\Omega_{RF}$ as well as the microwave electric field strength. Consequently, the measurement of the microwave electric field strength can be converted into the measurement of the spectral splitting interval, which is the idea behind conventional EIT-AT methods.This measurement is direct and traceable to fundamental physical constants $\hbar$, providing advantages like self-calibration and so on\cite{CL_Degen_2017_Quantum_sensing}. It is worth noting that the minimum resolvable interval of EIT-AT, which is constrained by the linewidth of EIT, determines the lower limit of the Rydberg atom electric field sensor based on the EIT-AT splitting method\cite{Holloway_2017,Zhou_2022}. Conventional EIT-AT methods have achieved a lower limit of approximately 3 mV/cm for microwave field detection\cite{Holloway_2017}. Therefore,overcoming the constraints imposed by the EIT linewidth and extending the measurement capability of the EIT-AT method to a weaker microwave fields has become a major research interest in this field. Enhancing EIT sensitivity for Rydberg atoms has been pursued through two dominant methods: the increase of EIT-AT splitting interval \cite{Simons_Matt_2016,Jia_2021_PRB,Yuan_2022,Xiubin_2021} and the decrease of EIT linewidth\cite{Kaiyu_2020,Zhou_2023,Fabian_2022}. Strategies employed to augment EIT-AT splitting interval include microwave field detuning\cite{Simons_Matt_2016}, auxiliary microwave fields \cite{Jia_2021_PRB,Yuan_2022}, and modulation dispersion of microwave amplitude\cite{Xiubin_2021}. Techniques employed to reduce the EIT linewidth encompass cold atom methodologies \cite{Kaiyu_2020,Zhou_2023} and three-photon excitation\cite{Fabian_2022}. While these methods show distinct advantages, their limitations have motivated researchers to explore more approaches.

Currently, extensive research has been conducted on EIT in the presence of magnetic field, providing insight for using magnetic field as a way to improve EIT-AT methods' measurement limits. Rydberg atoms exhibit a wide range of spectral information and potential applications when subjected to a magnetic field, which has attracted significant attention\cite{Bao_2017,Jia_2020_AO_2108-2113,Bao_2015,Bao_2016,Xue_2019,Ma_2017,Zhang_2018,Naber_2017,Cheng_2017,Su_2022_Optimizing,XU_2022_SAPB,Zi-Shan-Xu:73201,Jia_2020_AO_8253-8258,Jia_2021_PRB,Whiting_2016}. Shanxia Bao \textit{et al.}\cite{Bao_2017} and Fengdong Jia \textit{et al.}\cite{Jia_2020_AO_2108-2113} demonstrated the frequency locking of the coupling light using Rydberg EIT under magnetic field modulation. Shanxia Bao \textit{et al.}\cite{Bao_2015,Bao_2016}, Yongmei Xue \textit{et al.} \cite{Xue_2019}, L. Ma \textit{et al.}\cite{Ma_2017}, Linjie Zhang \textit{et al.} \cite{Zhang_2018}, and J. B. Naber \textit{et al.} \cite{Naber_2017} have investigated the linear and nonlinear variations of Rydberg energy levels under different static magnetic field strengths, along with the corresponding Rydberg EIT spectra. H. Cheng \textit{et al.}\cite{Cheng_2017}, H. J. Su \textit{et al.}\cite{Su_2022_Optimizing}, Z.S. Xua \textit{et al.}\cite{XU_2022_SAPB} and Zi-Shan Xu \textit{et al.}\cite{Zi-Shan-Xu:73201} have contributed to optimizing and improving the intensity and resolution of Rydberg EIT spectra using a static magnetic field. In another study, Fendong Jia \textit{et al.}\cite{Jia_2020_AO_8253-8258} enhanced the signal-to-noise ratio of the readout region by modulating the Rydberg EIT spectrum with an alternating magnetic field, and further combined it with an atomic mixer to achieve phase-sensitive detection\cite{Jia_2021_PRB}. Additionally, D. J. Whiting \textit{et al.}\cite{Whiting_2016} successfully measured the hyperfine structure constant of the excited state using EIT under a magnetic field. By combining the findings from the above research, it is anticipated that the EIT-AT spectrum will exhibit even more intricate structures under the combined influence of a magnetic field and a microwave electric field. This will provide valuable insights into understanding the alterations in level structures caused by external field effects, and potentially offer new insights for precise measurements related to spectral splitting intervals. However,no systematic study has been conducted on the EIT-AT splitting spectrum behavior of rubidium Rydberg atoms under a static magnetic field to the best of our knowledge.

In this paper, we theoretically and experimentally investigate the Rydberg EIT and EIT-AT splitting of $^{87}$Rb vapor under the combined influence of a magnetic field and a microwave field and further studied the enhancement of the Rydberg atom microwave field sensor performance by the static magnetic field.The structure of this paper is as follows: In Sec. \ref{Sec. 2}, we develop a simplified analytical formula for the EIT-AT spectrum under the static magnetic field and the microwave field following the model proposed by H. J. Su \textit{et al.}\cite{Su_2022_Optimizing}. In Sec. \ref{Sec. 3}, we introduce the experimental setup. In Sec. \ref{Sec. 4}, we present the experimental results of EIT and EIT-AT under magnetic field and compare them with the theoretical calculations provided in Sec. \ref{Sec. 2}. We also demonstrate the potential of the static magnetic field in enhancing the measurement of weak microwave fields using the EIT-AT splitting method. We concludes the paper in Sec. \ref{Sec. 5}.

\section{Theoretical Model} 
\label{Sec. 2}

\begin{figure}[ht!]
\centering\includegraphics[width=13cm]{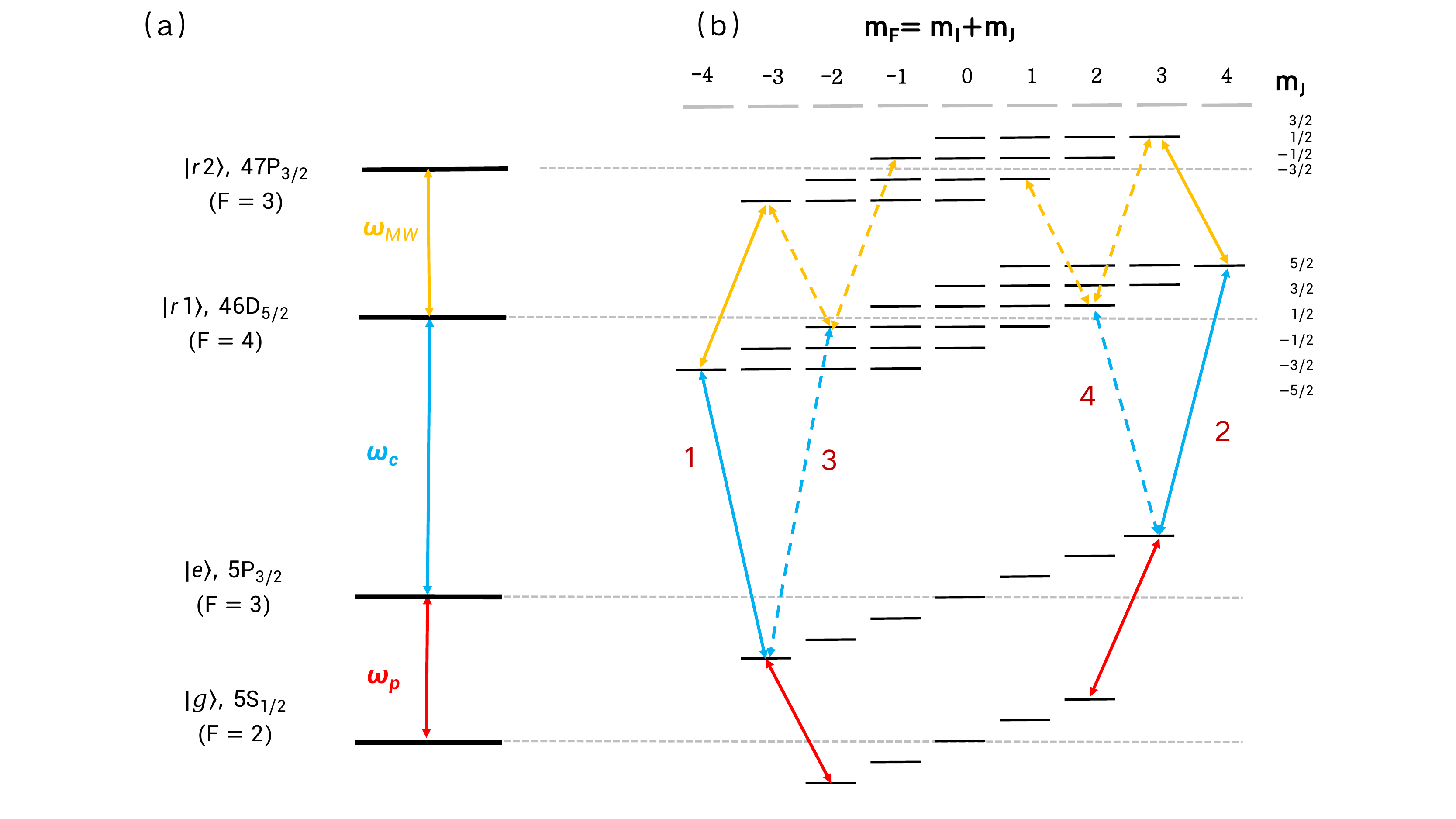}
\caption{(a)Degenerate four-level system energy level without magnetic field. (b)Non-degenerate energy level diagram of a four-level system under magnetic field. The arrows indicate possible transitions between different levels, with labels denoting distinct transition paths. The probe light is represented by the red, the coupling light by the blue, and the microwave field by the yellow.}
\label{Fig.1}
\end{figure}

In this study, we investigate the energy levels and optical fields as illustrated in Fig. \ref{Fig.1}.  The degenerate energy levels of the four-level system in the absence of a magnetic field is depicted in Fig. \ref{Fig.1}(a): ground state $|g> (5S_{1/2})$, excited state $|e> (5P_{3/2})$, two Rydberg states $|r_1> (46D_{5/2})$ and $|r_2> (47P_{3/2})$. The Zeeman splitting of the hyperfine structure in the presence of a magnetic field is shown in Fig. \ref{Fig.1}(b), where the degeneracy of the Zeeman sublevels is broken by the magnetic field. We choose the basis of F and $m_F$ to describe the system,where $F=I+J$ and $m_F=m_I+m_J$. $I$ is the nuclear spin angular momentum, which is 3/2 for $^{87}$Rb. The arrows in Fig. \ref{Fig.1}(b) indicate possible transitions between different levels, with labels denoting distinct transition paths.

We first consider the three-level system $|g>$, $|e>$, and $|r_1>$. The influence of the magnetic field on the EIT spectrum of this three-level system can be described by equating the Zeeman effect to the change in the optical field detuning\cite{Su_2022_Optimizing}. In Ref.\cite{Su_2022_Optimizing}, Hsuan-Jui Su \textit{et al.} solved the optical Bloch equations and the Maxwell-Schrodinger equation (MSE) in a ladder-type three-level Rb EIT model. They obtained an analytical solution for the transmittance of the probe light as a function of the detuning of the coupling light under the influence of the magnetic field\cite{Su_2022_Optimizing}:

\begin{equation}
T(\Delta_c) = exp[-\alpha\frac{\Gamma_e^2(4\gamma^2+4\delta^2)+2\gamma\Omega^2_c\Gamma_e}{(2\gamma\Gamma_e+\Omega_c^2-4\Delta_p\delta)^2+(4\Delta_p\gamma+2\Gamma_e\delta)^2}],
\end{equation}

where $\Delta_c$ represents the detuning of the coupling light, $\Omega_c$ is the optical density, $\gamma$ is the coherence dephasing rate between the ground state and Rydberg state, and $\Gamma_e$ is the linewidth of the excited state $|e> (5P_{3/2})$. By fitting the experimentally measured EIT spectrum, we can obtain a set of corresponding parameter values. According to the Winger-Eckart theorem, the coupling strength between hyperfine levels is determined by the magnetic dipole matrix elements. Thus the ratio of different transition paths are proportional to their corresponding magnetic dipole matrix elements.

In the presence of a magnetic field $B$, the total detuning is given by:
\begin{equation}
    \delta = B\times(\Delta_g-\Delta_e+\Delta_{r_1}+\Delta_D)+\Delta_c,
\end{equation}
\begin{equation}
    \Delta_g = \frac{\mu_B}{\hbar}g_{F,g}m_F,
\end{equation}
\begin{equation}
    \Delta_e = \frac{\mu_B}{\hbar}g_{F,e}m_{F'},
\end{equation}
\begin{equation}\label{eq:6}
    \Delta_{r_1} = \frac{\mu_B}{\hbar}g_{J,r_1}m_{r_1}.
\end{equation}
Here, $B$ represents the magnetic field strength. $\Delta_g,\Delta_e,\Delta_{r_1}$ are the Zeeman shift of the ground state, excited state, and the Rydberg state respectively. $g_{F,g},g_{F,e},g_{J,r_1}$ represent the Landé g-factors of the ground state $5S_{1/2}$, excited state $5P_{3/2}$, and the Rydberg state $46D_{5/2}$, which are 1/2, 2/3, and 6/5 in this study. The Doppler shift is simplified as the detuning: $\Delta_D = (\Delta_e-\Delta_g)\frac{\lambda_p-\lambda_c}{\lambda_c}$. In our experiment, we lock the frequency of the coupling light and scan the frequency of the probe light to obtain the EIT-AT splitting spectra. While the probe light detuning is much smaller than the Doppler broadening, there exists a relationship $\Delta_c = \frac{\omega_c}{\omega_p}\Delta_p$\cite{Mack_2011}. Therefore, we obtain the relation between the probe light transmittance and the probe light detuning: $\delta = B\times(\Delta_g-\Delta_e+\Delta_{r_1}+\Delta_D)+\frac{\omega_c}{\omega_p}\Delta_p$.

Based on the work in Ref. \cite{Su_2022_Optimizing} and the splitting behavior of EIT under a detuned microwave field\cite{Simons_Matt_2016}, we construct a simplified analytical formula for the EIT-AT spectrum under the combined influence of the static magnetic field and the microwave field. Specifically, due to the dressing effect of the microwave field, each $m_F$ state which corresponds to one EIT transparency peak will splits into two branches. We use detuning $\Delta_S$ to represent the splitting interval of the spectrum with respect to EIT and obtain the total detuning of the system as: $\delta = B\times(\Delta_g-\Delta_e+\Delta_{r_1}+\Delta_D)+\frac{\omega_c}{\omega_p}\Delta_p+\Delta_S$.

Energy levels will shift under magnetic field, causing the originally resonant microwave electric field to detune. This leads to an asymmetry in the EIT-AT splitting doublet compared to the EIT position\cite{Simons_Matt_2016}. Representing the smaller splitting interval by $\Delta_{S_1}$ and the larger one by $\Delta_{S_2}$, combining the the EIT-AT splitting interval expression in off-resonance conditions $\delta_{MW}=\sqrt{\delta^2_{MW}+\Omega^2_{MW}}$\cite{Simons_Matt_2016}, we are able to derive the EIT-AT splitting interval expression under both magnetic field and microwave electric field as $\Delta_{S_1}=\frac{1}{2}\cdot(\sqrt{\delta^2_{MW}+\Omega^2_{MW}}-\delta_{MW})$ and $\Delta_{S_2}=\frac{1}{2}\cdot(\sqrt{\delta^2_{MW}+\Omega^2_{MW}}+\delta_{MW})$, where $\Omega_{MW}$ is the resonant Rabi frequency of the microwave field, $\delta_{MW}$ represents the microwave detuning under the magnetic field, which can be calculated by Eq. (\ref{eq:6})-(8).

\begin{equation}\label{eq:7}
    \Delta_{r_2} = \frac{\mu_B}{\hbar}g_{J,r_2}m_{r_2},
\end{equation}
\begin{equation}\label{eq:8}
    \delta_{MW} = \Delta_{r_2}-\Delta_{r_1}.
\end{equation}

Combining the expression for the transition probability in Reference \cite{Tannoudji1992AtomphotonI}, we derive the strengths of the two peaks in the EIT-AT spectrum $f_1=\frac{1}{2}\cdot\frac{\Omega^2_{MW}}{\Omega^2_{MW}+\delta^2_{MW}}$ and $f_2=1-f_1$. By applying $\Delta_S$ to represent the EIT-AT splitting interval and adjusting the values of $\Omega_C$ for different transition paths, we can obtain the splitting behavior of EIT branches with involved $m_F$ states and quantitatively calculate their maximum splitting interval.

\section{Experimental Setup and Methods}\label{Sec. 3}

\begin{figure}[ht!]
\centering\includegraphics[width=13cm]{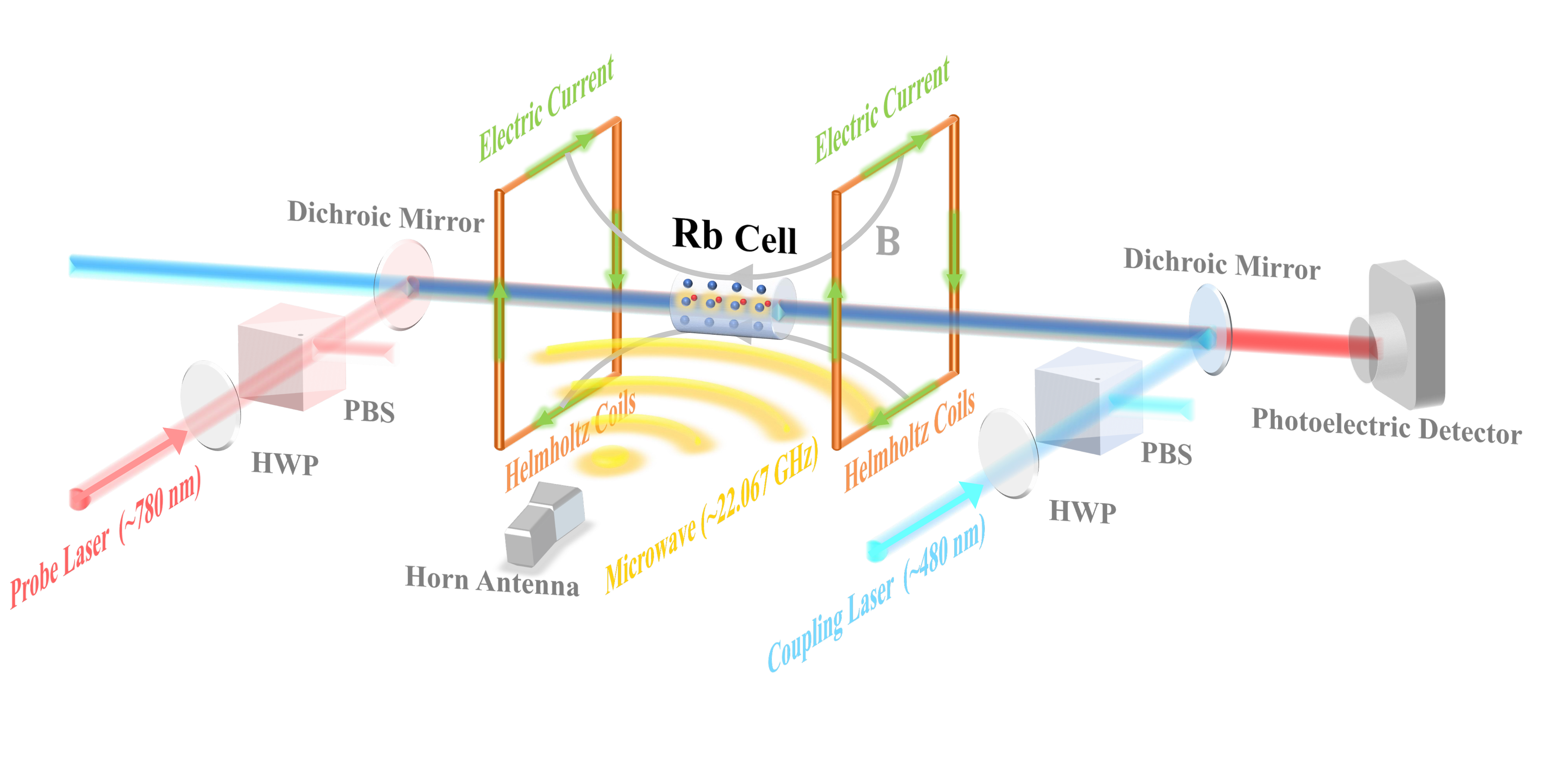}
\caption{Experimental setup. The $\thicksim$780 nm probe lights and $\thicksim$480nm coupling lights were set linearly polarized and collinearly counterpropagated through the Rb vapor cell. The static magnetic field was generated by a pair of rectangular Helmholtz coils (50 cm $\times$ 50 cm) placed perpendicular to the axis of the Rb cell and the microwave electric field was provided by a signal source (8340 B, HP Corporation) connected to a Horn Antenna placed by the side of the Rb cell. HWP:half wave plate; PBS: polarizing beam splitter.}
\label{Fig.2}
\end{figure}

The experimental setup for EIT and EIT-AT spectrum in a rubidium vapor cell under a magnetic field is shown in Fig. \ref{Fig.2}. The experiments were conducted under room temperature. The probe light, provided by a DL100 laser from Toptica, has a wavelength of $\thicksim$780 nm, and corresponds to the transition from $^{87}$Rb $5S_{1/2}$ to $5P_{3/2}$(shown as red arrows in Fig. \ref{Fig.1}(b)). The coupling light, provided by a TA-SHG-Pro laser from Toptica, has a wavelength of $\thicksim$480 nm and corresponds to the transition from $^{87}$Rb $5P_{3/2}$ to $46D_{5/2}$(shown as blue arrows in Fig. \ref{Fig.1}(b)). The probe and coupling lights were linearly polarized and collinearly counterpropagated through the Rb vapor cell. The spectroscopic signals of the interacting probe light were collected using a photodetector (Thorlabs, PDA36A2). A uniform DC magnetic field was provided by a pair of rectangular Helmholtz coils (50 cm $\times$ 50 cm) placed perpendicular to the axis of the Rb cell, ensuring a uniform static magnetic field near the vapor cell. The magnetic strength B at the location of the vapor cell was measured by a Gauss meter as a reference beforehand. The microwave electric field was provided by a signal source (8340 B, HP Corporation) connected to a Horn Antenna placed by the side of the Rb cell, with a frequency of 22.067233 GHz, corresponding to the transition from $^{87}$Rb $46D_{5/2}$ to $47P_{3/2}$ (shown as yellow arrows in Fig. \ref{Fig.1}(b)).

\section{Results and Discussion} \label{Sec. 4}
\subsection{Experimental spectra and theoretical analysis of Rydberg EIT splitting under static magnetic field}\label{4.1}

\begin{figure}[h]
\centering\includegraphics[width=13cm]{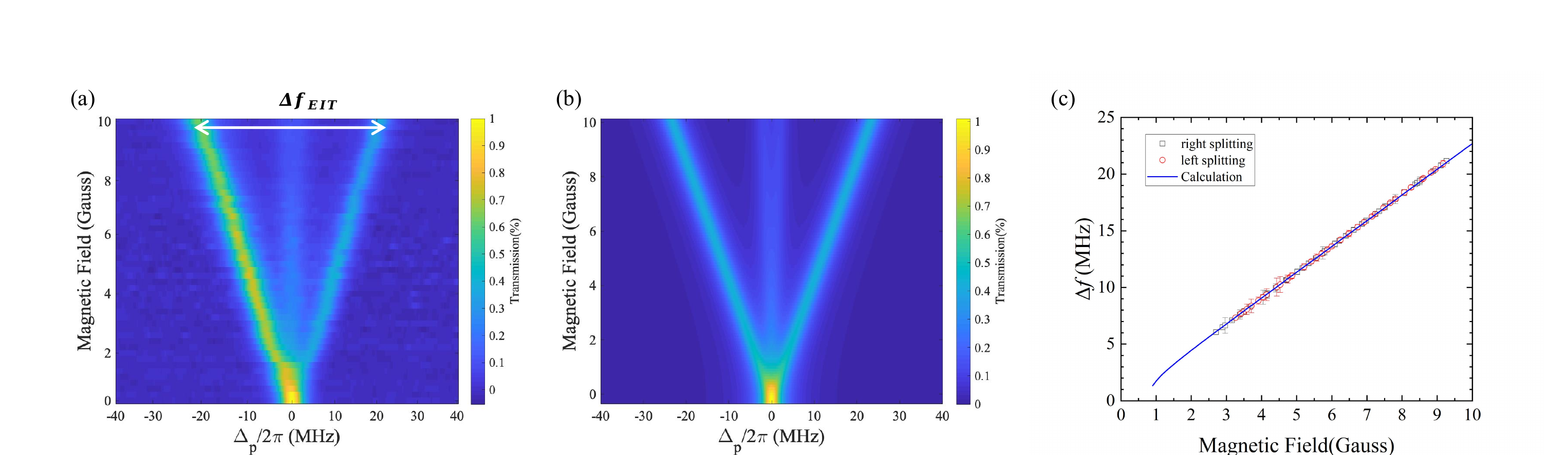}
\caption{(a) The experimental results of the EIT spectrum under a magnetic field of 0-10 G are as follows: The probe light has an optical power of approximately 60 $\mu W$ with a spot size of $\thicksim$400 $\mu m$ in radius, while the probe light has an optical power of approximately 40 mW with a spot size of $\thicksim$450 $\mu m$ in radius.The interval between the outmost EIT spectra are denoted as $\Delta f_{EIT}$ in Sec. \ref{4.3}. (b) The theoretical calculations of the EIT spectrum under a magnetic field of 0-10 G are as follows. The parameters used are: $\alpha=0.257$, $\gamma=2\pi\times3.3\times10^6$ MHz, and $\Omega_c=2\pi\times1.2\times10^6$ MHz (for CGC = 1). The two side peaks correspond to the transition paths 1 and 2 in Fig. \ref{Fig.2}(b) for CGC = 1, while the two middle peaks correspond to the transition paths 3 and 4 in Fig. \ref{Fig.2}(b) for CGC = 0.32. The theoretical calculations qualitatively agree well with the experimental results.(c) We calibrate the experimental results using theoretical calculations to determine the relationship between the EIT splitting interval and the magnetic field strength. The blue solid line represents the theoretical calculation results with the expression $\Delta f=(2.293\pm0.003)\cdot B-(0.173\pm0.021)$. The red lines indicate the splitting intervals of the left branch spectral lines in Fig. \ref{Fig.3}(b) after magnetic field calibration, while the black lines represent the splitting intervals of the right branch spectral lines after magnetic field calibration.}
\label{Fig.3}
\end{figure}

To investigate the magnetic field's influence on Rydberg EIT, we varied the current in Helmholtz coils to obtain a static magnetic field range of 0-10 G. We employed a frequency-locked coupling light while scanning the probe light within the range $\Delta_p=\pm 2\pi\times40$ MHz to acquire the spectrum of the $5S_{1/2}-5P_{3/2}-46D_{5/2}$ Rydberg transition under static magnetic field. The experimental spectrum is shown in Fig. \ref{Fig.3}(a). The experimental results revealed that the Rydberg EIT spectrum split symmetrically into four peaks as the magnetic field increased. The outer two peaks exhibited larger splitting intervals, indicating their higher sensitivity to the static magnetic field, while the inner two peaks displayed smaller splitting intervals, suggesting a lower sensitivity. These distinctions are due to the transitions of different Zeeman sublevels, which can be determined by referencing the energy level diagram in Fig. \ref{Fig.1} and the experimental results. Specifically, when subjected to a static magnetic field ranging from 0 to 10 G, Zeeman sublevels with the same absolute value of $m_F$ exhibited symmetric energy level shifts, leading to the appearance of symmetrical peaks in the EIT spectrum. Our theoretical analysis revealed that the two outermost peaks in Fig. \ref{Fig.3}(a) corresponded to transitions labeled as 1 and 2 in Fig. \ref{Fig.1} with CGC = 1, while the relatively magnetic-field-insensitive two middle peaks corresponded to transitions labeled as 3 and 4 in Fig. \ref{Fig.1} with CGC = 0.32.

By applying the theoretical model shown in Sec. \ref{Sec. 2}, we are able to fit and calculate the experimental EIT spectra obtained under the magnetic field.The results was shown in Fig. \ref{Fig.3}(b). By comparing Fig. \ref{Fig.3}(a) and (b), we can see a high level of agreement between experimental spectrum the theoretical calculations in terms of peak positions, thereby confirming the suitability of the energy level model. However, slight differences were observed in peak strengths and linewidths. The differences and explanations are as follows. Firstly, theoretical calculations predicted symmetric peak strengths for the peaks corresponding to excitation route with Zeeman sublevels of the same absolute value of $m_F$, while the experimental EIT spectra displayed asymmetric peak strengths. This difference is caused by the imperfect polarization of the probe and coupling light after passing through the atomic vapor cell. Secondly, the theoretical calculations yielded narrower linewidths compared to the experimental observations. This disagreement arose because the analytical solution in the theoretical calculations only considered the peak position variation caused by the Doppler effect, without accounting for the residual Doppler effect-induced broadening due to the dismatch of $\lambda_p$ and $\lambda_c$. Still, the theoretical calculations based on the analytical solution has accurately reproduced the peak positions, satisfying the requirements of focusing on the EIT-AT splitting interval. 

We perform a quantitative analysis of the spectral position information presented in Fig. \ref{Fig.3}(c) to accurately calibrate the magnetic field strength experenced by the atoms. Due to the limited EIT linewidth resolution observed in the experimental results, precise fitting of the peak positions for the two intermediate lines is not feasible. As a result, we calibrate the magnetic field faced by the atoms by utilizing the two outer lines, which correspond to the $m_F=\pm\,4$ states. By fitting the peak positions corresponding to the same $m_F$ state excitation configuration under the 0-10 G static magnetic field in the experiment, we establish the relationship between the experimental splitting interval of the two outer lines and the magnetic field, as depicted by the black hollow squares and red hollow circles in Fig. \ref{Fig.3}(c). These points exhibit a linear correlation with the magnitude of the magnetic field. Notably, when the magnetic field is less than 2.5 G, the splitting interval cannot be resolved because the magnetic field is smaller compared to the EIT linewidth. By performing a linear fit on the spacing between the outer two lines, we derive the following expression: $\Delta f=(2.293\pm0.003)\cdot B-(0.173\pm0.021)$, which enables the absolute calibration of the magnetic field. The blue solid line in Fig. \ref{Fig.3}(c) represents the results obtained using the calibrated magnetic field and the theoretical analytical solution for the three-level system, exhibiting good agreement with the experimental findings. It is worth noting that for small magnetic field values, the calculated spectral intervals cannot be distinguished, which is consistent with the experimental observations. Through the investigation of the $^{87}$Rb EIT splitting behavior under a magnetic field, we not only identify the magnetic sublevels and excitation channels involved in Rydberg EIT, but also provide an absolute calibration of the magnetic field experienced by the atoms. This serves as a solid foundation for further exploration of the EIT-AT splitting behavior under the influence of microwave fields.

\subsection{Experimental Spectra and theoretical analysis of Rydberg EIT-AT splitting under static magnetic field} \label{4.2}

\begin{figure}[h!]
\centering\includegraphics[width=13cm]{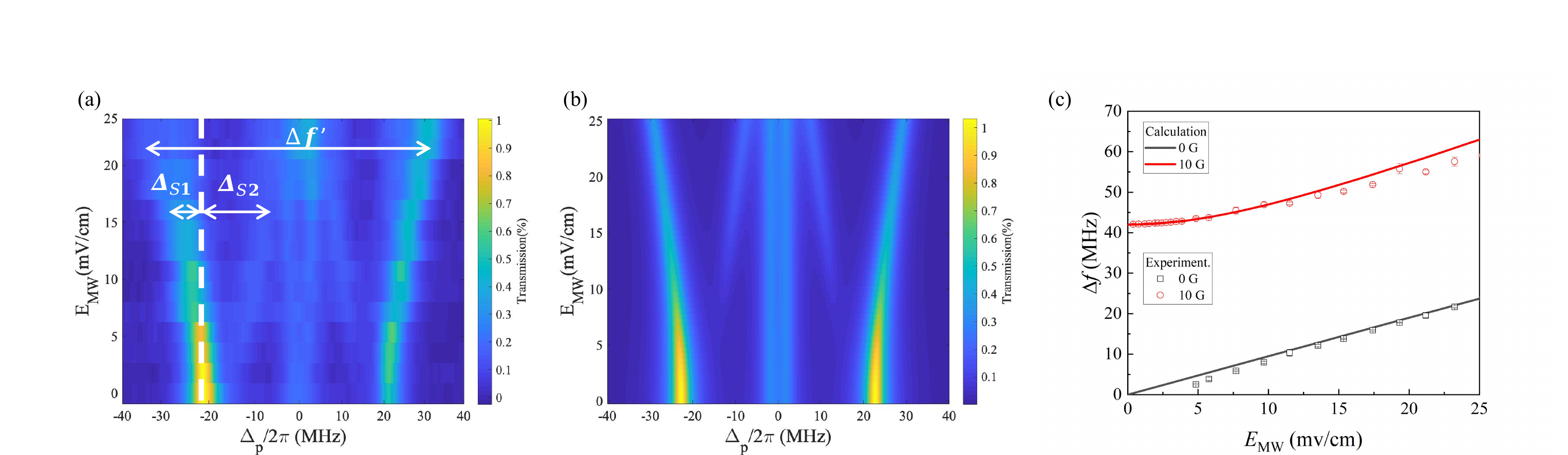}
\caption{(a)The experimental results of the EIT-AT split spectra under a static magnetic field of 10 G are presented as follows. For the experimental parameters, the probe light is characterized by a power of approximately 60 $\mu W$ and a beam size of approximately 400 $\mu m$ in radius. Similarly, the coupling light exhibits a power of approximately 40 mW and a beam size of approximately 450 $\mu m$ in radius.The interval between the asymmetry EIT-AT peak and the EIT peak is denoted as $\Delta_{S_1}$ and $\Delta_{S_2}$ respectively. The interval between the outmost EIT-AT spectra are defined as the new EIT-AT splitting interval $\Delta f'$. (b) Theoretical calculations of the EIT-AT split spectrum under the same 10 G static magnetic field are conducted with the following parameters: $\alpha=0.257$, $\gamma=2\pi\times3.3\times10^6$ MHz, and $\Omega_c=2\pi\times1.2\times10^6$ MHZ (for CGC = 1), and $\Omega_{MW}=2\pi\times1.35\times10^6$ MHz (for CGC = 1). The theoretical calculations are in qualitative agreement with the experimental results displayed in Fig. \ref{Fig.4}(a). (c)The relationship between the EIT-AT intervals and the magnitude of the microwave field for 0 and 10 G static magnetic fields. The black hollow squares correspond to the EIT-AT splitting interval as a function of the microwave field at 0 G magnetic field, while the black solid line represents the linear fit of the EIT-AT splitting interval at resonance under 0 G magnetic field. The red hollow circles depict the spacing between the outermost two EIT-AT spectral lines obtained experimentally, and the red solid line represents the calculated result of the EIT-AT splitting interval under 10 G by Eq. (\ref{eq:10}). The calculated curves demonstrates excellent agreement with the experimental data.}
\label{Fig.4}
\end{figure}

We conducted measurements on the Rydberg EIT-AT spectrum under a combination effect of static magnetic field and microwave field. The magnetic field can be fixed at 0 G to 10 G when measuring the EIT-AT spectrum with microwave field strength ranging from 0 to 25 mV/cm. Fig. \ref{Fig.4}(a) illustrates the EIT-AT split behavior under a static magnetic field at 10 G with different microwave field strengths. It is important to point out that we maintained the microwave field frequency the same as the resonance frequency with the Rb atoms at zero magnetic field throughout the experiment. The results in Fig. \ref{Fig.4}(a) demonstrate that after the degeneracy is broken by the magnetic field, each EIT spectrum splits as the microwave field strength increases.The outermost EIT spectral lines exhibit asymmetrical EIT-AT splitting in terms of their positions and strengths. The position asymmetry arises from the magnetic field-induced shifting of Zeeman sublevels with nonzero $m_F$, causing the microwave field frequency to deviate from resonance with the Rydberg level. This behavior is consistent with the previously reported characteristics of Rydberg EIT-AT splitting under non-resonant microwave field frequencies\cite{Simons_Matt_2016}. In contrast, the two middle spectral lines experience insignificant splitting due to the insensitivity of their excited states to the magnetic field and weak or zero Rydberg-pair states interactions\cite{Walker_2008PRA_blockage}. Consequently, no further theoretical analysis was conducted for these lines. At approximately 25 mV/cm, an increase in spectral intensity was observed due to the overlap of the splitting EIT lines. 

We qualitatively calculated the EIT-AT spectrum shown in Fig. \ref{Fig.4}(a) based on the analytical solutions for the four-level system, which agrees with the experimental results well. This calculation further validates the selection of energy levels and the applicability of the chosen model. Similarly, when calculating the resonant frequency of the microwave field, we took into account the proportions between different hyperfine structure levels of the atoms at different Rydberg states. The calculated ratio of $m_F$ states corresponding to the outermost and innermost EIT lines is approximately 1.35:1. A narrower linewidth was acquired by theoretical calculation compared to the experimental result. The reason is that since the two middle lines are insensitive to both the magnetic and microwave fields, we only considered the splitting of the largest mF state, i.e.the two outermost EIT spectral lines in the theoretical calculation. The experimental results in Fig. \ref{Fig.3}(a) may involve additional transition paths, leading to a broader linewidth compared to the theoretical calculation. Nevertheless, this discrepancy already satisfies our requirements regarding the peak positions and splitting intervals of EIT-AT. Besides, the theoretical calculation indicates that the strengths of the spectral lines corresponding to the excited routes with the same absolute value of $m_F$ are symmetric while that are asymmetric observed in experiments. The reason is the same as discussed in Section \ref{4.1}. Through our investigation of rubidium atoms' EIT-AT behavior under static magnetic field, we discovered that the interval of the outermost EIT spectra are no longer limited by the EIT linewidth. Therefore, the EIT-AT splitting can still be resolved under weak microwave fields, which provides insights for enhancing the EIT-AT splitting interval method in RF electric field measurement.

\subsection{Enhancing the lower measurement limit utilizing EIT-AT splitting of the maximum $m_F$ state in a static magnetic field}\label{4.3}
Now we try to measure the strength of the microwave field using the outermost spectral intervals, as depicted in Fig. \ref{Fig.4} (a) and (b). There are three motivations behind this choice. Firstly, the outermost spectral lines demonstrate the highest intensity when subjected to EIT optical pumping. Secondly, the overlapping of intermediate spectral lines poses challenges in accurately determining and reading peak positions, particularly under strong microwave field. Lastly and most importantly, our focus in this study is on the expanding of EIT-AT intervals induced by static magnetic field. And the outermost spectral intervals indicate the possibility that Zeeman effect-induced EIT-AT splitting in the magnetic field can enlarge the EIT-AT intervals, thereby enabling the measurement of weaker microwave field strengths. By fitting the positions of the two outermost peaks in Fig. \ref{Fig.4}(a), we establish the relationship between the experimental splitting interval and the microwave field, as shown by the red empty circles in Fig. \ref{Fig.4}(c). For comparison, we also provide the relationship between the EIT-AT splitting interval and the microwave field at zero magnetic field, illustrated by the black empty squares in Fig. \ref{Fig.4}(c). Notably, a linear relationship between the EIT-AT splitting interval and the microwave field is observed when $E > 10 mV/cm$ at zero magnetic field. This suggests that we can determine the Rabi frequency of the microwave field and achieve absolute calibration of its strength in this range. However, for microwave fields below 10 mV/cm, the relationship between the EIT-AT splitting interval and the microwave field deviates from linearity, and for E < 5 mV/cm, the EIT-AT splitting becomes indistinguishable, reaching the lower limit of microwave field strength measurement using the conventional EIT-AT splitting method.When a magnetic field of 10 G is applied, the redefined splitting interval $\Delta f'$, i.e. the interval between the outermost two spectral lines, persists and exhibits a monotonic relationship with the microwave field, as shown in Fig. \ref{Fig.4}(c). Notably, we discovered that the minimum measurable microwave field is 0.38 mV/cm, which extends the measurement limit by a factor of 12.6.

In the following analysis, we aim to develop an analytical formula that describes the relationship between the redefined EIT-AT splitting interval $\Delta f'$ under the magnetic field and the microwave field. Notably, the two outermost EIT lines in Fig. \ref{Fig.4}(a) exhibit an asymmetric EIT-AT splitting in terms of position and intensity, consistent with the behavior of Rydberg EIT-AT splitting reported in Ref. \cite{Simons_Matt_2016} when the microwave field frequency is off-resonance. Building upon the findings in Ref. \cite{Simons_Matt_2016}, we establish a new analytical solution to describe the EIT-AT splitting interval exhibited by peaks with identical $\lvert m_F \rvert$ values under the magnetic field. Ref. \cite{principle_of_laser_cooling} provides a relationship between the EIT-AT splitting interval and the detuning:

\begin{equation}\label{eq:9}
    \Delta f=\sqrt{\delta^2_{MW}+\Delta f^2_0},
\end{equation}

where $\Delta f$ is the detuning, $\Delta f_0$ represents the EIT-AT splitting interval at resonance. The strong peak of the EIT-AT splitting is closer to the EIT peak without microwave field. We denote this smaller interval as $\Delta_{S_1}=\frac{1}{2}\cdot(\sqrt{\delta^2_{MW}+(\beta\cdot\Delta f_0)^2}-\delta_{MW})$, where $\beta$ denotes the ratio of mF states corresponding to the outer and inner EIT spectral lines.This assumption is reasonable since the formula reduces to the resonance case when $\delta_{MW}$ approaches 0, and the microwave field no longer interacts with the Rydberg energy levels as $\delta_{MW}$ approaches infinity, resulting in no EIT splitting, and the formula returns to Eq. (\ref{eq:9}). Similarly, the interval between the strong peak of the right EIT-AT splitting and its corresponding EIT peak is denoted as $\Delta_{S_2}=\frac{1}{2}\cdot(\sqrt{\delta^2_{MW}+(\beta\cdot\Delta f_0)^2}-\delta_{MW})$. Additionally, we consider the interval caused by the magnetic field in zero microwave field as $\Delta f_0$. Then the interval between the outermost two spectrum has the expression:

\begin{equation}\label{eq:10}
    \Delta f=\sqrt{\delta^2_{MW}+(\beta\cdot\Delta f_0)^2}-\delta_{MW}+\Delta f_{EIT},
\end{equation}

where $\delta_{MW}$ is the detuning and can be calculated by Eq.(6)-(8), $\Delta f_0$ represents the EIT-AT splitting interval at resonance. The calculated $\beta$ value is $\beta\approx1.35$, indicating a favorable ratio constant greater than 1 for EIT-AT splitting measurements. $\Delta f_{EIT}$ is the splitting interval under a magnetic field, as depicted in Fig. \ref{Fig.3}(b).
    
Note that $\Delta f_0$ can be obtained by linear fitting of the EIT-AT splitting interval at 0 G, as shown by the solid black line in Fig. \ref{Fig.4}(c). Using Eq. (\ref{eq:10}), the calculated results for the outermost two spectral lines at 10 G are shown by the solid red line in Fig. \ref{Fig.4}(c). The calculated results demonstrate a good agreement with the experimental data. This confirms our experimental objective: the EIT-AT splitting induced by the Zeeman effect under a magnetic field can increase the EIT-AT interval, thereby characterizing weaker microwave electric field strengths. Moreover, this interval can be described by a simple analytical formula.

\section{Conclusion} \label{Sec. 5}
In this study, we investigate the splitting behavior of $^{87}$Rb Rydberg atoms’ EIT and EIT-AT spectrum in a static magnetic field. By introducing an equivalent detuning to describe the magnetic field effect, we derive a simple analytical solution that accurately explains the experimental observations. Under the influence of the magnetic field, the EIT peaks exhibit multiple spectral lines, with symmetric splitting observed in the same $m_F$ state. The calculated results are in excellent agreement with the experimental data. Through the analysis of the EIT splitting behavior of rubidium atoms in a magnetic field, we not only identify the magnetic sublevels and excitation transition routes involved in EIT system, but also achieve absolute calibration of the magnetic field experienced by the atoms. Moreover, we investigate the split spectrum of EIT-AT under the combined effect of static magnetic field and microwave field. Our findings reveal a monotonic relationship between the interval of the two outermost peaks of EIT-AT splitting (corresponding to the largest absolute $m_F$ values) and the microwave field at 10 G. Interestingly, we determine that the minimum measurable microwave field is 0.38 mV/cm, which represents a 12.6-fold improvement in the lower limit of microwave field measurement. More interestingly, we propose a simple analytical formula to explain the relationship between the splitting interval of these two spectral lines and the electric field strength. The calculated results exhibit good agreement with the experimental data, thereby providing a simplified approach for investigating EIT-AT splitting in the presence of a magnetic field. Our work deepens our understanding of the Rydberg EIT-AT spectrum under the combined influence of a static magnetic field and a microwave field, and unveils its potential applications. We demonstrate how the spectral splitting of different $m_F$ states can enhance the lower limit of electric field measurement, which is helpful to precise microwave electric field measurements in weak fields regime.

\begin{backmatter}
\bmsection{Funding}
Beijing Natural Science Foundation under Grant No. 1212014, the National Natural Science Foundation of China (Grant Nos. 11604334, 11604177 and U2031125), the Strategic Priority Research Program (grant no. XDB28000000) of the Chinese Academy of Sciences, the Fundamental Research Funds for the Central Universities, and National Key Research and Development Program of China (Grant Nos. 2017YFA0304900 and 2017YFA0402300).

\bmsection{Acknowledgments}
We are grateful for the fruitful discussions with Professor T F Gallagher at University of Virginia;Dr. Yuechun Jiao, Dr. Linjie Zhang, and Prof. Jie Ma at Shanxi University; Dr. Kaiyu Liao and Prof. Hui Yan at South China Normal University; and Dr. Fei Meng at the National Institute of Metrology, China.

\bmsection{Disclosures}
The authors declare no conflicts of interest.

\bmsection{Data availability}
Data underlying the results presented in this paper are not publicly available at this time but may be obtained from the authors upon reasonable request.

\end{backmatter}

\bibliography{EIT-inB-ref}

\end{document}